\documentclass[twoside,11pt]{article}

\usepackage{graphicx}

\usepackage{amssymb}
\usepackage{amsmath}
\usepackage{amsfonts}
\usepackage{wasysym}
\usepackage{revsymb}
\usepackage{amsbsy}
\usepackage{bm}

\usepackage{titlesec}

\usepackage{cite}

\titleformat*{\section}{\bfseries}
\titleformat*{\subsection}{\bfseries}

\begin{document}           

\title{\Large 
On primordial matter production induced by spatial curvature in the early universe}
\author{V.E. Kuzmichev, V.V. Kuzmichev\\[0.5cm]
\itshape Bogolyubov Institute for Theoretical Physics,\\
\itshape National Academy of Sciences of Ukraine, Kyiv, 03143 Ukraine}

\date{}

\maketitle

\begin{abstract}
\noindent
In this note, it is shown that nonvanishing spatial curvature produces primordial matter in the initially empty universe due to quantum gravity effects. This matter decays faster than radiation and is described by a stiff equation of state. The quantum Hamiltonian constraint equation for the universe with the maximally symmetric geometry is solved in the semi-classical approximation. The extra energy density and pressure of quantum origin that appear in the generalized Friedmann equations describe primordial matter and modify the expansion history of the early universe.
\end{abstract}


\section{Introduction}
\label{sec:1}

According to the standard cosmological model, the universe went through a stage of inflation after the Big Bang, during which its size increased exponentially, 
so that the universe began to look like an expanding homogeneous and isotropic emptiness \cite{RR,DZS,S80,G81,AS82,GP82,BST83}.
In order to fill the universe with matter and radiation at a very early stage, 
it is necessary to supplement the model with a physical mechanism allowing the production of particles.
The transfer of energy from a slowly moving classical inflaton field to standard matter particles at the end of the accelerated expansion period was 
initially analyzed using perturbation theory.  It was assumed that the inflaton field was coupled to standard matter through the appropriate interaction 
Lagrangian. There was a subsequent shift from simple single-field inflation models to more sophisticated ones, including those that 
treat matter fields quantum mechanically or consider the quantum production of matter particles in a classical background (see, for example, 
Refs.~\cite{K96,KKLT98,ABC10,McD20}).

In most approaches, the effect of spatial curvature on the evolution of the universe is not taken into account, since the very early universe ($\sim 10^{-35}$ s) 
is assumed to be spatially flat, with a high degree of precision ($\sim 10^{-50}$) concerning departure from flatness \cite{RR}. These estimations 
stem from the classical equations of general relativity, although the possibility of deviation from zero spatial curvature is not completely ruled out 
\cite{VMS19,H19,ACMV23}. In the realm of the quantum theory the nonvanishing spatial curvature can appear 
due to quantum fluctuations of geometry (for recent discussions on the subject, see, e.g., Refs.~\cite{CCM17,C19,CMP20,BM25}), 
which in turn can become a source of matter production. Since inflation takes place at scales that are a few orders of magnitude below the Planck scale, 
quantum gravitational effects may come into play \cite{KTV19}. The same applies to the early post-inflation era.

The investigations of particle creation in expanding universes, carried out in the theory of quantized fields on curved spacetimes, including cosmological
Robertson--Walker spacetimes, are well-known \cite{P68,P69,P71,BD82,PT09}. In that treatment, the gravitational field itself is not quantized and its 
curvature is predetermined. 

In the present note, we attempt to consider the matter production in a semiclassical approximation to quantum cosmology. Unlike studying this process within 
the framework of quantum field theory in curved space, in the approach proposed here, the newly created matter has a reverse effect on 
geometry, modifying it.
We show that quantum gravity effects can cause nonvanishing spatial curvature to produce matter described by a stiff equation of state. 
Its energy density $\rho$ decreases as $\rho \sim a^{-6}$, where $a$ is a scale factor. Such a stiff matter decreases faster than 
radiation, but the impact of the stiff matter created in the early stages of the universe's evolution does not seem to be enough to explain the Hubble tension 
\cite{KR22,V21,V23,K24}.

In modern physics, a stiff matter is not something exotic.
In the scope of general relativity, the cosmological model, in which the very early universe was supposed to be filled with a gas of cold baryons 
with a stiff equation of state, has been first introduced in Refs.~\cite{Zel62,Zel72} at the phenomenological level.
It is believed that the stiff matter era preceded the radiation era, the dust matter era, and the dark energy era. This kind of matter appears in numerous 
cosmological models. The stiff matter arises in the case, when the universe is filled with a cosmological scalar field whose energy density is dominated by 
its kinetic term. Certain cosmological models, where dark matter is made of relativistic self-gravitating Bose--Einstein condensates (BECs), feature the stiff 
matter \cite{Cha15,Cha15a}. The effects of a stiff pre-recombination era caused by various mechanisms (e.g., employing the axion kination approach) influence 
the energy spectrum of the primordial gravitational waves \cite{Oik23,Co20,Oik23a}. Another example is the origin of stiff matter in a two scalar field model that 
incorporates non-Riemannian measures of integration and involves an additional $R^{2}$--term as well as scalar matter field potentials of appropriate form so 
that the action is invariant under global Weyl-scale symmetry \cite{Gue23,Gue14}. 

\section{Classical theory}
\label{sec:2}

Consider the homogeneous and isotropic cosmological system (universe), whose geometry is determined by the Robertson--Walker line element, 
\begin{equation}\label{0}
ds^{2} = N^{2} d \tau^{2} - a^{2} (\tau) d \Omega_{3}^{2},
\end{equation}
where $N$ is the lapse function whose choice is arbitrary \cite{ADM62}, $a$ is the cosmic scale factor, and $d \Omega_{3}^{2}$ is the line-element of a constant curvature space with curvature parameter $\kappa = +1, 0, -1$.

The dynamical behaviour of such a universe with a perfect fluid distribution of matter is described by the Friedmann equations, which can be written as \cite{Lan75}
\begin{equation}\label{1}
\left(\frac{d a}{d \eta} \right)^{2} = \frac{8 \pi G}{3 c^{4}} \rho a^{4} - \kappa a^{2},
\end{equation}
\begin{equation}\label{1.1}
\frac{d^{2} a}{d \eta^{2}}  = \frac{4 \pi G}{3 c^{4}} a^{3} \left(\rho - 3 p \right) - \kappa a,
\end{equation}
where the evolution of the scale factor $a$ is given in conformal time $\eta$, which is related to proper time $t$ 
by the differential relation $d t = N d \tau = a d \eta$. The energy density $\rho$ and pressure $p$ of a perfect fluid are measured in GeV cm$^{-3}$,
while the scale factor $a$ and proper time $t$ are measured in cm. The conformal time $\eta$ is dimensionless\footnote{During an arc-time interval 
$\Delta \eta$, a light signal travels $\Delta \eta$ radians around the universe.}. The pressure is given by
\begin{equation}\label{1.2}
p = - \frac{a}{3} \frac{d \rho}{d a} - \rho.
\end{equation}

It is convenient to switch to dimensionless units. In the following, the scale factor and proper time will be measured in Planck length units 
$l_{p} = \sqrt{G \hbar/(c^{3})}$, and the energy density will be taken in Planck energy density units $\rho_{p} = 3 c^{4}/(8 \pi G l_{p}^{2})$.

We will assume that the universe is originally filled with matter with the energy density 
$\rho_{m}$ and radiation with the energy density $\rho_{\gamma}$, so that $\rho = \rho_{m}  + \rho_{\gamma}$. 

Equations (\ref{1}) and (\ref{1.1}) can be rewritten in the compact dimensionless form
\begin{equation}\label{1.3}
\left(\frac{d a}{d \eta} \right)^{2} = 2 a M(a) + E - \kappa a^{2},
\end{equation}
\begin{equation}\label{1.4}
\frac{d^{2} a}{d \eta^{2}} = M(a) + a \frac{d M(a)}{d a} - \kappa a,
\end{equation}
where it is denoted 
\begin{equation}\label{1.5}
2 M(a) = \rho_{m} a^{3}, \quad E = \rho_{\gamma} a^{4}.
\end{equation}
Let us note that in dimensional physical units, $M(a)$ has the dimension [Energy], and $E$ has the dimension [Energy × Length].

The set of equations (\ref{1.3}) and (\ref{1.4}) will be compared with the equations of the quantum theory below.

For an empty expanding universe, Eq.~(\ref{1.3}) simplifies,
\begin{equation}\label{2}
\frac{d a}{d \eta} = \sqrt{- \kappa}\, a \quad \mbox{or} \quad \frac{d a}{d t} = \sqrt{- \kappa}.
\end{equation}
Given the normalization choice of $a (\eta = 0) = 1$ and the initial condition of $a (t = 0) = 0$, we have the solutions to these equations
\begin{equation}\label{3}
a = e^{\sqrt{- \kappa}\, \eta} \quad \mbox{and} \quad a = \sqrt{- \kappa}\, t.
\end{equation}
We see that in conformal time $\eta$, the spatially open ($\kappa = -1$) universe expands exponentially, whereas the spatially closed ($\kappa = 1$) 
universe oscillates. In terms of proper time $t$, the empty universe expands according to the linear law for $\kappa = -1$ or is described in complex 
quantities for $\kappa = 1$. This circumstance is confirmed by a formal replacement $a^{2} \rightarrow - a^{2}$ when passing from a space with positive 
curvature to a space with negative curvature \cite{Lan75}. A spatially flat empty universe is static, $a = const$. In the framework of general relativity, in the 
empty universe even at $\kappa \neq 0$, there is no source of production of matter. Such a source is revealed after the transition from general relativity to 
quantum theory.

\section{Quantum theory}
\label{sec:3}

There are numerous variants of the formulation of the quantum gravity equations in the literature\footnote{At present there are up to twenty alternative 
approaches to quantum gravity (see, e.g., Ref.~\cite{E11} and references therein).}. We will use the approach involving constrained canonical 
quantization in geometrodynamical variables. Following Dirac's scheme of canonical quantization \cite{D58} for the  
universe with the maximally symmetric geometry, the basic equation can be reduced to the form 
\begin{equation}\label{5}
\left(- \partial_{a}^{2} + \kappa a^{2} - 2 a H_{\phi} - E \right) \Psi = 0,
\end{equation}
where $\Psi = \Psi (a, \phi)$ is a state vector, $H_{\phi}$ is the Hamiltonian of the matter field $\phi$, which by definition fills the universe from the very 
beginning along with a perfect fluid used as a `reference system' enabling us to
recognize the instants of time and points of space (see, e.g., Refs.~\cite{Ku91,BM96}) and taken in the form of relativistic matter (radiation). 
All other notations are as in Eq.~(\ref{1.3}). 

Equation (\ref{5}) is the quantum version of the Hamiltonian constraint equation. It can be considered as an analog of the Wheeler--DeWitt equation \cite{W68,DW62,DW67}. After averaging over a complete orthonormal set of states $| \Phi_{i} \rangle$ of the field $\phi$, which diagonalizes 
the Hamiltonian $H_{\phi}$, Eq.~(\ref{5}) reduces to \cite{K08,K13}
\begin{equation}\label{7}
\left(- \partial_{a}^{2} + \kappa a^{2} - 2 a M_{i}(a) - E \right) \psi_{i} (a) = 0,
\end{equation}
where it was taken into account that $\langle \Phi_{i} | H_{\phi} | \Phi_{k} \rangle = M_{i}(a) \delta_{i k}$ and denoted 
$\psi_{i} (a) \equiv \langle \Phi_{i} | \Psi \rangle$. The quantity $M_{i}(a)$ is the mass of matter in the universe in the discrete and/or continuous 
$i$th state. The subscript $i$ is inactive and can be omitted below.

In solving this equation, we will proceed in a way that resembles the Madelung--Bohm formalism which allows one to transform the 
quantum-mechanical equations into hydrodynamic equations \cite{Ma27,B52,So91,He15,He16,K25}.

The solution of Eq.~(\ref{7}) will be found in the polar form
\begin{equation}\label{8}
\psi (a) = A(a)\,e^{i S(a)},
\end{equation}
where the amplitude $A(a)$ and the phase $S(a)$ are real functions. Substituting Eq.~(\ref{8}) into Eq.~(\ref{7}) gives two equations
\begin{equation}\label{9}
\partial_{a} (A^{2} \partial_{a} S) = 0,
\end{equation}
\begin{equation}\label{10}
\left(\partial_{a} S \right)^{2} + \kappa a^{2} - 2 a M(a) - E - Q = 0,
\end{equation}
where
\begin{equation}\label{10.1}
Q = \frac{\partial_{a}^{2} A}{A}
\end{equation}
is an analog of the quantum Bohm potential \cite{B52}. From Eq.~(\ref{9}), it follows 
$A = const / \sqrt{\partial_{a} S}$. Then
\begin{equation}\label{11}
Q = \frac{3}{4} \left(\frac{\partial_{a}^{2} S}{\partial_{a} S} \right)^{2} - \frac{1}{2}\, \frac{\partial_{a}^{3} S}{\partial_{a} S}.
\end{equation}
The set of Eqs.~(\ref{9}) -- (\ref{10}) is exact and strictly equivalent to Eq.~(\ref{7}). In hydrodynamic interpretation, Eq.~(\ref{9}) can be considered as a 
continuity equation. It represents the conservation law of current density $|\psi|^{2} \partial_{a} S$ for a  ``fluid'' of density $|\psi|^{2} = A^{2}$.
Equation (\ref{10}) plays a role similar to that of the law of conservation of energy in fluid dynamics. 

In the classical approximation, the quantum potential $Q$ is neglected ($Q \sim l_{p}^{4} \sim \hbar^{2}$ in dimensional units \cite{K13,B52}) and
Eq.~(\ref{10}) turns into the Hamilton-Jacobi equation, whose solution is the ``principal function'' of Hamilton (action functional) $S$. For the momentum 
$\pi_{a} = \partial_{a} S$ conjugate to the variable $a$, from the minisuperspace action, it follows $\pi_{a} = - \frac{d a}{d \eta}$ 
(see, for example, Ref.~\cite{Kie25}).

Following Bohm's interpretation \cite{B52}, Eq.~(\ref{10}) can still be regarded as the Hamilton-Jacobi equation even when $\hbar \neq 0$ and
$- \partial_{a} S = \frac{d a}{d \eta}$ can still be regarded as the velocity which characterizes the expansion or contraction of the universe 
as the motion in minisuperspace. The quantum potential $Q$ (\ref{10.1}) extends the classical dynamics into the quantum regime.

In this case, in the semiclassical limit, Eq.~(\ref{10}) takes the following form,
\begin{equation}\label{11.1}
\left(\frac{d a}{d \eta} \right)^{2} = 2 a M(a) + E - \kappa a^{2} + Q.
\end{equation}
Differentiating Eq.~(\ref{11.1}) with respect to $\eta$, we obtain
\begin{equation}\label{11.2}
\frac{d^{2} a}{d \eta^{2}} = M(a) + a \frac{d M(a)}{d a} - \kappa a + \frac{1}{2} \frac{d Q}{d a}.
\end{equation}
Comparing Eqs.~(\ref{11.1}) and (\ref{11.2}) with Eqs.~(\ref{1.3}) and (\ref{1.4}), we see that taking quantum effects into account has led to the appearance of 
additional terms on the right-hand side of the Friedmann equations. To clarify the physical meaning of these terms, let us rewrite these equations in standard 
form expressed through energy densities and pressures.

Using the expression for pressure (\ref{1.2}) and the definitions (\ref{1.5}), we get
\begin{equation}\label{11.3}
\left(\frac{d a}{d \eta} \right)^{2} = a^{4} \rho_{tot} - \kappa a^{2},
\end{equation}
\begin{equation}\label{11.4}
\frac{d^{2} a}{d \eta^{2}} = \frac{1}{2} a^{3} \left(\rho_{tot}  - 3 p_{tot}  \right) - \kappa a,
\end{equation}
where
\begin{equation}\label{11.5}
\rho_{tot} = \rho_{m} + \rho_{\gamma} + \rho_{q}, \quad p_{tot} = p_{m} + p_{\gamma} + p_{q}
\end{equation}
with
\begin{equation}\label{11.6}
\rho_{q} = \frac{Q}{a^{4}}, \quad p_{q} = \frac{1}{3 a^{3}} \left(- \frac{d Q}{d a} + \frac{Q}{a} \right).
\end{equation}
The extra terms brought in by the quantum potential $Q$ can be interpreted as contributions from an additional matter component with the energy 
density $\rho_{q}$ and pressure $p_{q}$. The continuity equation for the expanding (or contracting) universe can be written as
\begin{equation}\label{11.7}
\frac{d \rho_{tot}}{d \eta} + \frac{3}{a} \frac{d a}{d \eta} \left(\rho_{tot} + p_{tot} \right) = 0.
\end{equation}

Explicit expressions for $Q$ for different cases, when one or another component of matter dominates in the 
universe, are given in Ref.~\cite{K25}. Let us consider the special case of an empty universe with spatial curvature ($\kappa \neq 0$, $M = 0$, $E =0$). 
Then Eq.~(\ref{10}) reduces to
\begin{equation}\label{12}
\left(\partial_{a} S \right)^{2} = - \kappa a^{2} + Q.
\end{equation}
In the dimensional units, 
the potential $Q \sim \hbar^{2}$ \cite{K13} and therefore the nonlinear equation (\ref{12}) can be solved using the WKB approximation.
Keeping only zero-order terms of the expansion of $(\partial_{a} S)^{2}$ in a power series in $\hbar^{2}$, we obtain the quantum potential (\ref{11}) 
for this system,
\begin{equation}\label{13}
Q = \frac{3}{4} \frac{1}{a^{2}}, 
\end{equation}
so that the quantum additions to the energy density and pressure (\ref{11.6}) have the form
\begin{equation}\label{14}
\rho_{q} = \frac{3}{4} \frac{1}{a^{6}}, \quad p_{q} = \rho_{q}.
\end{equation}
The energy density $\rho_{q}$ (\ref{14}) obtained in the semiclassical approximation turned out to be the same for the spatially closed and open universes. 

According to Eq.~(\ref{11.1}), in the case of the empty universe with spatial curvature and quantum addition, the generalized Friedmann equation can be represented as
\begin{equation}\label{16}
\left(\frac{d a}{d \eta} \right)^{2} = - \kappa a^{2} + \frac{3}{4} \frac{1}{a^{2}} \quad \mbox{or} \quad \left(\frac{d a}{d t} \right)^{2} = - \kappa + \frac{3}{4} \frac{1}{a^{4}}.
\end{equation}

Let us find the solutions of these equations with the initial condition $a(0) = 0$.
For a spatially closed universe with $\kappa = 1$, we have
\begin{equation}\label{19}
a (\eta) = \left(\frac{\sqrt{3}}{2} \sin(2 \eta) \right)^{1/2}.
\end{equation}
Taking into account the relation $d t = a d \eta$, we can express the solution in parametric form for the proper time $t$,
\begin{equation}\label{20}
t = \frac{3^{1/4}}{\sqrt{2}} \int_{0}^{\eta (t)}\! d \eta \sqrt{\sin (2 \eta)} = 
\frac{3^{1/4}}{\sqrt{2}} \left[E \left( \frac{\pi}{4}, \sqrt{2} \right) - E \left(\frac{\pi}{4} - \eta, \sqrt{2} \right) \right],
\end{equation}
where $E$ is the incomplete elliptic integral of the second kind.

For a spatially open universe with $\kappa = - 1$, the solution can be written as
\begin{equation}\label{21}
a (\eta) = \left(\frac{\sqrt{3}}{2} \sinh(2 \eta) \right)^{1/2},
\end{equation}
or in parametric form  for the proper time $t$
\begin{equation}\label{22}
\begin{split}
t & = \frac{3^{1/4}}{\sqrt{2}} \int_{0}^{\eta (t)}\! d \eta \sqrt{\sinh (2 \eta)} = \frac{3^{1/4}}{\sqrt{2}} \left[\frac{u \sqrt{u^{4} + 1}}{u^{2} + 1} -
\frac{1}{2} F \left(\arccos \frac{u^{2} - 1}{u^{2} +1}, \frac{1}{\sqrt{2}} \right) \right.\\
& \left. + E \left(\arccos \frac{u^{2} - 1}{u^{2} +1}, \frac{1}{\sqrt{2}} \right) +
\frac{1}{2} F \left(\pi, \frac{1}{\sqrt{2}}  \right) - E \left(\pi, \frac{1}{\sqrt{2}}  \right) \right], \quad u^{4} \equiv \frac{4}{3} a^{4},
\end{split}
\end{equation}
where $F$ is the incomplete elliptic integral of the first kind.

The curvature term becomes negligible in the expansion rate at early times. By dropping the term with the curvature parameter $\kappa$ in (\ref{16}), 
we obtain
\begin{equation}\label{15}
a \approx \left(\sqrt{3} \,\eta \right)^{1/2} \quad \mbox{or} \quad a \approx \left(\frac{3}{2} \sqrt{3}\, t \right)^{1/3}.
\end{equation}
The dependence of the scale factor on time turns out to be exactly as it should be for a universe dominated by stiff matter with the equation of state 
$p_{q} = \rho_{q}$.

\section{Discussion}
\label{sec:4}

In order to determine properties of matter which corresponds to the energy density $\rho_{q}$ (\ref{14}), we rewrite the generalized Friedmann equation 
(\ref{16}) in dimensional units
\begin{equation}\label{17}
\left(\frac{d a}{d \eta} \right)^{2} = \frac{8 \pi G}{3 c^{4}} \rho_{q} a^{4} - \kappa a^{2},
\end{equation}
where
\begin{equation}\label{18}
\rho_{q} = \frac{9}{32 \pi}\frac{G}{c^{4}} \frac{(\hbar c)^{2}}{a^{6}}.
\end{equation}
The dependences of the energy density (\ref{18}) on $G$ and $\hbar$ illustrate its origin from quantum gravitational effects. 

In the approach under study, the constituents of cosmological fluid with the energy density $\rho_{q}$ are not specified and, basically, it can be linked to 
different types of matter. Let us consider several examples. One example is a fluid made of self-interacting BECs \cite{Cha15,Cha15a}. 
Actually, taking into account that the rest-mass density 
$\rho_{0} \sim a^{-3}$, we have $\rho_{q} \sim a^{-6} \sim \rho_{0}^{2}$ and $p_{q} = \rho_{q} = K \rho_{0}^{2}$, where the proportionality coefficient 
$K \sim \hbar^{2}$ as in the cosmology of a BEC fluid.

Another example comes from a comparison with the cosmological model containing matter with intrinsic half-integer spin. 
The inverse dependence of $\rho_{q}$ on the square of volume $\sim a^{-6}$ allows spin effects to be singled out.
In order to consider the contribution of spin effects to the energy density, it is necessary to use the theory of gravity which extends general relativity to include 
the spin of matter. The Einstein--Cartan theory is a theory in which the antisymmetric part of the affine connection coefficients (torsion) becomes an 
independent dynamic variable which can be associated with the spin density of matter (the quantum-mechanical spin of microscopic particles) 
\cite{Ki61,Sc62,He76,G86,Po10}.
On a macroscopic level, the contribution of particles with spin can be averaged and described in terms of Weyssenhoff spin fluid \cite{W47}. If the spins 
are randomly oriented, only the terms that are quadratic in the spin tensor do not vanish after averaging.

Comparing the generalized Friedmann equations containing quantum corrections with the Einstein--Cartan equations for a spin fluid, we find that they are equivalent if we identify the quantum additions to the energy density and pressure with the effective energy density and the effective pressure of a spin fluid,
\begin{equation}\label{19.2}
\rho_{q} =  \tilde{\rho} - \frac{2 \pi G}{c^{4}} s^{2}, \quad p_{q} =  \tilde{p} - \frac{2 \pi G}{c^{4}} s^{2},
\end{equation}
where $s^{2} = \frac{1}{2} \langle s_{\mu \nu} s^{\mu \nu} \rangle$ is the square of the spin density, $\tilde{\rho}$ and $\tilde{p}$ are the thermodynamic energy 
density and pressure of a stiff fluid with the equation of state $\tilde{p} = \tilde{\rho}$.

Introducing the particle number density $n$, the square of the spin density for a fluid can be written as
\begin{equation}\label{20.2}
s^{2} = \frac{1}{8} \left(\hbar c n \right)^{2}.
\end{equation}
The particle number density is inversely proportional to the volume, $n \sim a^{-3}$, and the energy density can be freely redistributed between both 
summands $\sim a^{-6}$ in Eq.~(\ref{19.2}). The square of the spin density (\ref{20.2}) applies to a fluid consisting of fermions with no spin polarization (see,
e.g., Ref.~\cite{Po10}). 

\section{Concluding remarks}
\label{sec:5}

In this note, we have demonstrated through direct calculations, without appealing to additional hypotheses, that a matter component with a stiff equation of state emerges in an initially empty universe with nonvanishing spatial curvature due to quantum effects. The semiclassical approximation to the (Wheeler--DeWitt) quantum gravity equation is used. 

The dependence $\rho \sim a^{-6}$ of the energy density component generated by the quantum Bohm potential is universal \cite{K24}. Regardless of which 
matter component dominates in the universe (with the exception of radiation, for which the specified mechanism does not work), the quantum potential in the 
semiclassical approximation gives rise to a quantum addition to the energy density, which decays faster than radiation. 
In the approach proposed here, the source of stiff matter production in the universe is not directly related to physical matter, which is 
defined by the stress-energy tensor in Einstein's equations. On the contrary, summands that play a key role here are constructed from metric and describe 
geometry. The stiff matter production is provided by spatial curvature, which may deviate from zero due to quantum fluctuations. 
The stiff matter in an expanding universe arises precisely at the right stage, i.e., after the quantum era and before the radiation era ($\rho \sim a^{-4}$), non-relativistic matter era ($\rho \sim a^{-3}$), or the cosmological constant era ($\rho = const$). 

It is commonly assumed that, after inflation, the universe is radiation dominated.
But the cosmological history may have been more complicated with deviations from standard radiation domination occurring in the earliest epochs, before Big Bang nucleosynthesis.  The stiff matter era discussed in this note is one of such deviations
from radiation domination in the early universe. Recently, several proposals have been made regarding various topics, such as the generation of dark matter, matter-antimatter asymmetry, gravitational waves, and baryogenesis, among others, during a nonstandard expansion phase \cite{Aet21}. If the universe is dominated by the energy density with the equation of state stiffer than radiation, then dark matter perturbations grow faster than they do during radiation domination \cite{RTE18}. The stiff matter era can cause the universe to expand rapidly without altering the nucleosynthesis process. One interesting feature is that the electroweak baryogenesis scenario differs from radiation-dominated models \cite{OO17}. The presence of stiff matter affects the primordial element abundances and gravitational 
particle creation in the very early universe \cite{KT90,JP98,DS10,OM11,LV17}. The strongest constraints on a stiff matter come from the primordial $^{4}$He 
abundance (for the measurement of the primordial $^{4}$He abundance see, for example, Refs.~\cite{IT10,ITG14,A10,A21,M22}).
The impact of the stiff matter in the early stage of universe's evolution influences many physical processes, and it is worth being taken into account.

\section*{Acknowledgements}
This work was partially supported by The National Academy of
Sciences of Ukraine (Projects No.~0121U109612 and  No.~0122U000886) and by a grant from Simons Foundation International SFI-PD-Ukraine-00014573, PI LB.

\end{document}